\documentclass{article}

\usepackage{PRIMEarxiv}

\usepackage[utf8]{inputenc} 
\usepackage[T1]{fontenc}    
\usepackage{hyperref}       
\usepackage{url}            
\usepackage{booktabs}       
\usepackage{amsfonts}       
\usepackage{nicefrac}       
\usepackage{microtype}      
\usepackage{lipsum}
\usepackage{fancyhdr}       
\usepackage{graphicx}       
\graphicspath{{media/}}     

\usepackage{subcaption}
\usepackage{microtype}
\usepackage{todonotes}
\usepackage{amsmath}
\usepackage{amssymb}
\usepackage{listings}
\usepackage{mathtools}
\usepackage{array}
\usepackage{booktabs}
\usepackage{bm}

\newcommand{\R}{\mathrm  I\!\mathrm R}
\newcommand{\vF}{\mathbf{F}}
\newcommand{\vdelta}{\bm{\delta}}
\newcommand{\vc}{\mathbf{c}}

\newcommand{\vi}{\mathbf{i}}

\newcommand{\vn}{\mathbf{n}}
\newcommand{\vm}{\mathbf{m}}

\newcommand{\vp}{\mathbf{p}}
\newcommand{\vr}{\mathbf{r}}
\newcommand{\vs}{\mathbf{s}}
\newcommand{\vv}{\mathbf{v}}
\newcommand{\vx}{\mathbf{x}}

\newcommand{\mD}{\mathbf{D}}
\newcommand{\mE}{\mathbf{E}}
\newcommand{\mH}{\mathbf{H}}
\newcommand{\mI}{\mathbf{I}}
\newcommand{\mJ}{\mathbf{J}}

\newcommand{\cD}{\mathcal{D}}
\newcommand{\cH}{\mathcal{H}}
\newcommand{\cI}{\mathcal{I}}
\newcommand{\cM}{\mathcal{M}}
\newcommand{\cP}{\mathcal{P}}

\newcommand{\cS}{\mathcal{S}}

\newcommand{\diff}{\mathrm{d}}
\DeclareMathOperator*{\argmin}{arg\,min}
\DeclareMathOperator*{\argmax}{arg\,max}
\newcommand{\vNull}{\mathbf{0}}

\title{InverseVis: Revealing the Hidden with Curved Sphere Tracing
}

\author{
  Kai Lawonn \\
  Friedrich Schiller University Jena \\
  Jena, Germany\\
  \texttt{kai.lawonn@uni-jena.de} \\
  \And
  Monique Meuschke \\
  Otto-von-Guericke University Magdeburg \\
  Magdeburg, Germany\\
  \texttt{meuschke@isg.cs.uni-magdeburg.de} \\
  \And
  Tobias G{\"u}nther \\
  Friedrich-Alexander-Universit{\"a}t Erlangen-N{\"u}rnberg \\
  Erlangen, Germany\\
  \texttt{tobias.guenther@fau.de} \\
}

\begin{document}
\maketitle

\begin{abstract}
Exploratory analysis of scalar fields on surface meshes presents significant challenges in identifying and visualizing important regions, particularly on the surface's backside. 
Previous visualization methods achieved only a limited visibility of significant features, i.e., regions with high or low scalar values, during interactive exploration. In response to this, we propose a novel technique, \emph{InverseVis}, which leverages curved sphere tracing and uses the otherwise unused space to enhance visibility.
Our approach combines direct and indirect rendering, allowing camera rays to wrap around the surface and reveal information from the backside. 
To achieve this, we formulate an energy term that guides the image synthesis in previously unused space, highlighting the most important regions of the backside. 
By quantifying the amount of visible important features, we optimize the camera position to maximize the visibility of the scalar field on both the front and backsides. 
InverseVis is benchmarked against state-of-the-art methods and a derived technique, showcasing its effectiveness in revealing essential features and outperforming existing approaches.
\end{abstract}


\section{Introduction}
The visualization of scalar fields on three-dimensional surfaces is a critical aspect of data analysis and interpretation in various scientific and engineering domains. In fields like fluid dynamics, material science, and geophysics, understanding the intricate details of scalar fields such as temperature, pressure, or stress distribution is crucial. 
For example, scalar fields such as Wall Shear Stress (WSS) and pressure, when encoded on a 3D surface, 
significantly aid in the decision-making process in material science or in treating conditions like aneurysms or vessel stenosis. A visual representation provides an intuitive understanding of critical vessel wall regions, thereby supporting more informed and precise treatment strategies. 

\begin{figure}[t]%
{
  \centering
  \includegraphics[width=0.9\linewidth]{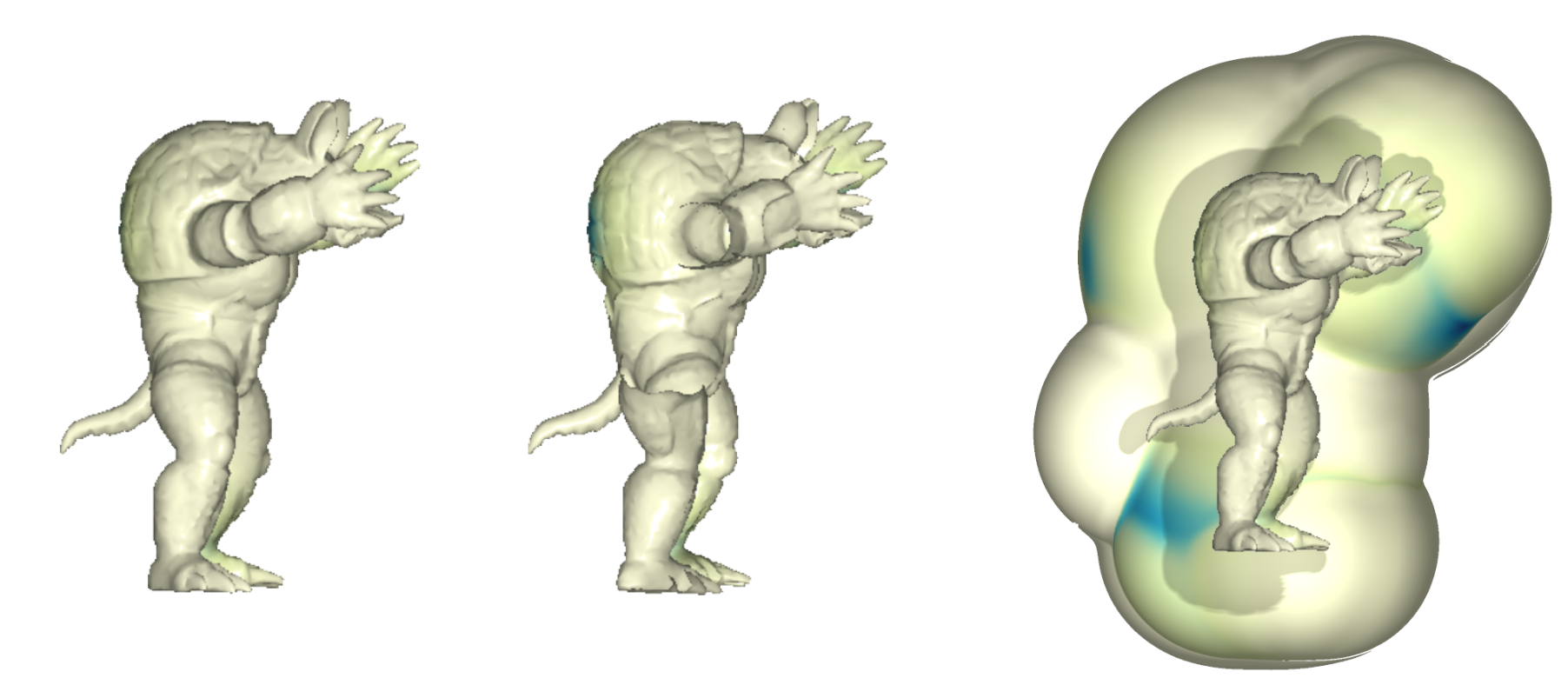}\\%
}
    \hspace{2.1cm}%
    (a) Front faces
     \hspace{2.1cm}%
    (b) Back faces
    \hspace{3.3cm}%
     (c) InverseVis 
    \caption{The first two images show the front and back faces. Hot spots cannot be detected while InverseVis detects them.}%
    \label{fig:InverseVisVsBackFront}%
\end{figure}

Color encoding a scalar field on a 3D surface is a widely used technique in data visualization. However, this approach presents several challenges for users when exploring 3D data, especially in the context of complex surfaces characterized by, e.g., irregular shapes, a high level of detail, uneven curvatures, or overlapping structures. When the user rotates the object, they must remember distinct surface regions. 
Moreover, certain parts of the surface, such as highly concave regions, can be difficult to see. Consequently, important data color-coded on these obscured areas may be overlooked, even when toggling front and back face rendering, as shown in Fig.~\ref{fig:InverseVisVsBackFront}, or they require extensive model manipulation to become visible, which can also hardly be overcome by animations~\cite{preim2020survey}.

Previous work applied projections~\cite{Neugebauer09,goubergrits2012statistical} or parametrizations~\cite{meuschke2019visual,eulzer2019temporal} to generate occlusion-free 2D visualizations of the encoded scalar fields. However, the resulting 2D map-based depictions are affected by distortion and leading to a misrepresentation of 
spatial relationships. Moreover, the transition from 3D to 2D inherently results in the loss of depth cues, which complicates the accurate interpretation of 3D structure and spatial relationships. Further, surfaces with complex topologies, such as those with holes or intricate curves, can be particularly challenging to parametrize. Often, the user has to define appropriate cut lines, where the parametrization can result in overlapping regions in 2D space, which can obscure important details and can be confusing. 
Another problem arises from the high cognitive load needed to relate the 2D representations to the original 3D structure.

\begin{figure}[t]%
{
 \centering%
 \includegraphics[width=\linewidth]{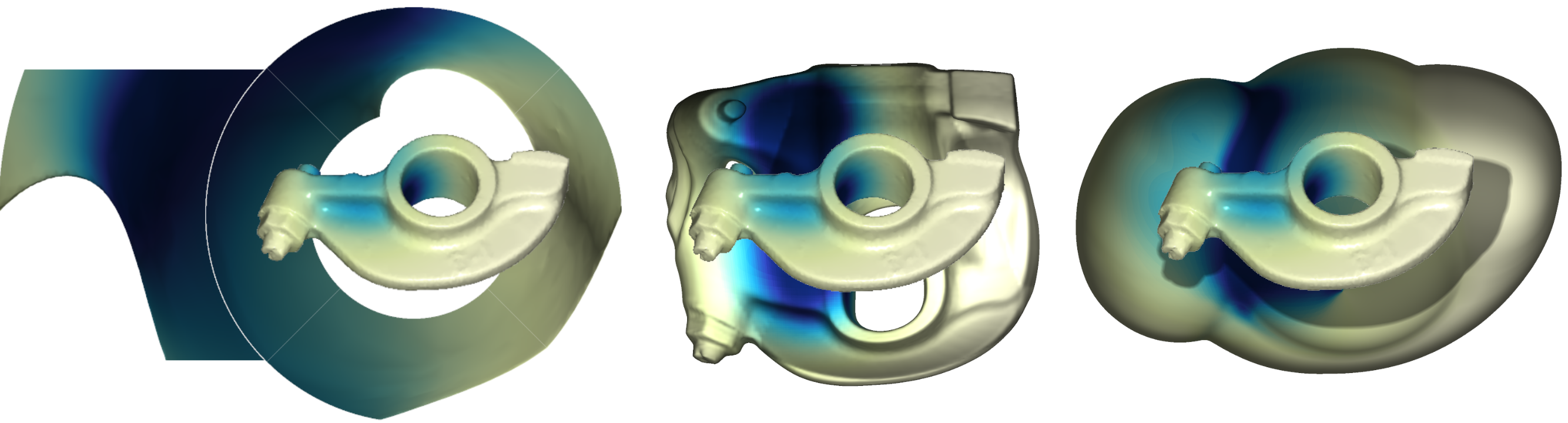}\\%
}
 \hspace{1.5cm}Neugebauer et al.~\cite{Neugebauer09}%
 \hspace{1.5cm}Quadratic optimized mirror%
 \hspace{2.3cm}InverseVis%
 \caption{The indirect visualization of scalar fields on the backsides of the rocker arm 
 is challenging. We propose a novel visualization method that utilizes the empty space around the surfaces to display the hidden parts of the scalar field with less occlusion and distortion.
 }%
\label{fig:teaser}%
\end{figure}

For this reason, we propose \emph{InverseVis}, a novel optimization-based surface visualization method that bends rays around the object to reveal the scalar field on the backside.
Similarly, Falk et al.~\cite{Falk07:Panorama} calculated force fields that bend rays to generate panorama maps of landscapes, which, however introduces deformations for all pixels and is calculated for a given viewpoint.
In our approach, the pixels that directly see the surface remain unchanged to support normal 3D perception.
To keep the visualization of the backside close to the geometry, we define an image space area that is located \emph{behind} the object, from which non-linear rays are released that fall in a gravitational potential towards the surface, thereby landing on the backside.
The scalar value found is then encoded at the location where the non-linear ray started.
To judge how much information is conveyed, 
we determine both directly and indirectly visible surface parts and integrate the displayed scalar field.
A higher score means that more information is shown.
Based on this score, we perform a gradient-based optimization of our method's parameter and identify an optimal viewpoint that conveys the most information.
We incorporate shadowing and surface shading to improve depth perception and compare our approach with a previous projection method and the optimization of a quadratic mirror surface behind the object, see Fig~\ref{fig:teaser}.
In summary, 
\begin{itemize}
    \item we propose a novel visual encoding technique that displays occluded scalar fields in otherwise empty pixels,
    \item we present a physically-inspired model that determines a mapping from the image space to the backside of a surface,
    \item we introduce a quality metric that measures the amount of directly and indirectly seen information,
    \item we perform an optimization of the viewpoint from which the most information is seen.
\end{itemize}

\section{Related Work}
The visualization of scalar fields on 3D surfaces is challenging due to occlusion. 
Existing methods that augment or alter the spatial arrangement can be sorted into two categories, as summarized next.
Later, we compare our approach with a method from each category.

\subsection{2D Map-based Visualizations}
Map-based visualizations are adept at representing and analyzing spatial data distributions on complex surfaces, negating the need for 3D rotation. They are often used to examine vessel wall pathologies like aneurysms or stenosis, displaying scalar fields such as WSS, thickness, and plaque distributions. 
2D maps are created through mesh projection or parameterization. Projection casts the 3D mesh onto shapes like cylinders or spheres, providing rapid computations but lacking bijective topology. Mesh parameterization 
allows for a direct bijective mapping between 3D surfaces and 2D maps. Both methods face challenges in handling branching surfaces, preserving features, standardizing layouts, and minimizing distortions.

\noindent
\textbf{Mesh Projections.} Neugebauer et al.~\cite{Neugebauer09} describe the generation of an overview map for the analysis of scalar data on cerebral aneurysm surfaces. The map is generated using a cube-map-based approach, where an axis-aligned cube is positioned at the center of the aneurysm, and the aneurysm's surface is projected onto five cube sides. 
This projection results in various map zones that encapsulate the 3D model, forming a ring. However, the projection introduces non-conformal spherical distortions and cut edges, which can cause orientation changes and confusion during interaction. Although a correlation tool aids in matching points on the aneurysm surface, the presence of projection artifacts and the complexity of the map's interactivity can hinder the clarity of the data and pose a learning challenge for users. Moreover, it does not prioritize critical surface structures, such as regions with high scalar values. 
Goubergrits et al.~\cite{goubergrits2012statistical} developed a method for mapping aneurysms to analyze WSS distributions, involving the repositioning of surface vertices towards the center of mass, forming a unit sphere, and employing azimuthal equidistant projection to preserve angular information. This technique is effective for convex structures but may lead to distortions in irregularly shaped aneurysms where the center of mass is external to the surface.
In heart disease research, particularly for analyzing left ventricular function, the Cardiac Bull’s Eye Plot (BEP) has been popular~\cite{kuehnel2006new,oeltze2006integrated,koehler20152d}.
This abstract 2D representation simplifies complex cardiac data by projecting the myocardium into 17 circularly arranged regions. Despite its usefulness, the BEP is optimized for myocardium grading. Due to this shape dependency, it can hardly be transferred to other 3D models.

\noindent
\textbf{Mesh Parametrizations.} Various mesh parametrization techniques have been developed, aiding in scalar field analysis. Eulzer et al.~\cite{eulzer2022vessel} presented an extensive overview of mapping vascular structures into 2D, and Kreiser et al.~\cite{kreiser2018survey} explored mapping techniques in other medical areas like the brain and colon. In the following, we review vessel parametrization techniques, as they were also employed in the validation of our approach.
Meuschke et al. employed Least Squares Conformal Maps (LSCM)~\cite{meuschke2017combined,meuschke2018exploration} and Spectral Conformal Parameterization (SCP)~\cite{meuschke2019visual} for angle-preserving aneurysm maps, and later As-Rigid-As-Possible (ARAP)~\cite{meuschke2021skyscraper} for area-preserving maps. These methods facilitate the visualization of multiple scalar fields but require advanced interaction techniques to maintain a correspondence between the 3D surface and the 2D map.
Antiga et al.~\cite{antiga2003automated,antiga2004robust} and Choi et al.~\cite{choi2017conformal,choi2020area} used conformal techniques for mapping vessel bifurcations, crucial to analyze vascular calcium deposits on the inside of the vessel wall. 
Eulzer et al.~\cite{eulzer2019temporal} and Karim et al.~\cite{karim2014surface} focused on the mitral valve and left atrium, respectively, using methods like those described by Lichtenberg et al.~\cite{lichtenberg2020mitral} for flattening complex structures.
Ma et al.~\cite{ma2012cardiac}, Nu{\~{n}}ez-Garcia et al.~\cite{nunezgarcia2019standard}, Paun et al.~\cite{paun2017patient}, and Roney et al.~\cite{roney2019universal} presented various techniques for mapping the ventricles and atria. These include quasi-conformal mapping, and methods using Laplace's equation.
Nevertheless, in contrast to our method, these aforementioned techniques fall short in providing spatial context surrounding the focal object. Moreover, often different additional inputs such as manually selected cut lines or a centerline are needed to generate the map. These key distinctions are why we do not draw direct comparisons between our approach and these parametrization methods. 

\subsection{Mirrors}
Virtual mirrors 
simulate a mirror-like interface, enabling users to view parts of a three-dimensional object that are typically obscured or difficult to access.
Navab et al.~\cite{Navab:2007} introduced the concept 
in augmented reality (AR), particularly for medical applications, such as navigated surgery. The virtual mirror is controllable within the AR environment and reflects the virtual part of the scene, adding a second perspective. 
Based on this work, Bichlmeier et al.~\cite{Bichlmeier:2006:tangible,bichlmeier2007laparoscopic,Bichlmeier:2009} evaluated the virtual mirror in advanced AR medical applications, such as navigated drilling in spine surgery and minimally invasive tumor resection in liver surgery. The virtual mirror is attached to tracked objects like surgical instruments, allowing it to reflect virtual entities of the AR scene. This integration simplifies its use, as there is no need for additional devices to control the mirror's position. The mirror reflects only virtual objects, which are registered with and superimposed on the real objects in the scene, like the anatomy of interest or a drilling device. It aids in planning and controlling surgical procedures by providing a view of regions that are not directly visible. 
However, the virtual mirror also presents certain disadvantages. Its use is limited to convex structures where the center of mass lies inside the surface, and it may produce area distortions in maps of irregularly shaped objects. The complexity of its interactive nature can pose a learning challenge for users. Moreover, while the mirror enhances visualization, it does not prioritize critical surface structures or formulate optimization problems for mapping important regions, e.g., with high scalar values.

\section{Problem Statement and Terminology}

\begin{figure}[t]%
    \centering%
    \includegraphics[width=0.75\linewidth]{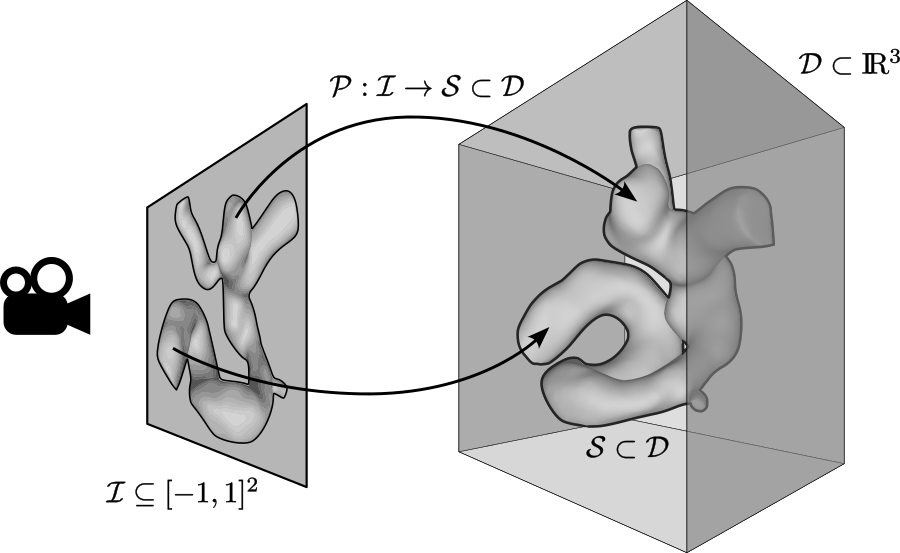}%
    \caption{Our goal is to find a mapping $\cP: \cI \rightarrow \cS$ that maps from the image plane $\cI$ to the surface domain $\cS$, for example, to show as much of the surface as possible or special areas with scalar fields.}%
    \label{fig:projection-overview}%
\end{figure}%

Occlusion of relevant information is among the most challenging problems in the visualization of three-dimensional spatial data.
When the spatial domain contains a set of surface geometries, the interesting question arises if the two-dimensional image plane can be mapped to the two-dimensional surface space, such that:
\begin{enumerate}
    \item object parts that are visible to the viewer remain visible,
    \item object parts that were occluded are displayed in blank areas,
    \item and emphasis is put onto relevant parts of the surface.
\end{enumerate}

In the remainder of this paper, we refer to the image domain as $\cI \subseteq [-1,1]^2$, which identifies a location in screen space in normalized device coordinates (NDC).
Further, a set of surfaces occupies the spatial domain $\cS \subset \cD$, which is a subset of the full spatial domain also known as world space $\cD \subseteq \mathbb{R}^3$.
Each location in the image domain $\vi\in [-1,1]^2$ corresponds to a world space location $\vi_\cD\in\cI_\cD$ on the near plane $\cI_\cD\subset\mathbb{R}^3$.
A camera model provides for each point on the near plane $\vi_\cD\in\cD$ a view ray direction $\vr_{\vi_\cD}$. 
When using an orthographic projection, the ray direction $\vr_{\vi_\cD}$ is the same for every point $\vi_\cD\in\cD$.
Under perspective projection, the view ray direction varies spatially.
Our goal is to find a partial map $\cP: \cI \rightharpoonup \cS$ that maps from a given image pixel to a point on the surface.
A schematic illustration is provided in Fig.~\ref{fig:projection-overview}.
Next, we introduce a common notation in which we describe methods that we later compare  with.
Afterwards, we introduce our novel optimization-based formulation.

\paragraph*{Visibility.}
Under a projective mapping, some parts of the image will see a surface while others do not.
We partition the image plane into three regions, depending on the visibility of the surfaces $\cS$:
\begin{align}
    \cI = \underbrace{\cI^+}_{\textrm{directly visible}} 
    \cup \underbrace{\cI^-}_{\textrm{indirectly visible}} 
    \cup \underbrace{\cI^0}_{\textrm{not visible}}
\end{align}
Region $\cI^+$ contains all pixels that directly see a surface, 
\begin{align}
    \cI^{+}= \{ \vi\in\cI \;|\; \min_{\lambda\ge 0, \vs\in\cS} \Vert \vi_\cD+\lambda\vr_{\vi_\cD}-\vs\Vert=0 \}.
\end{align}
Region $\cI^-$ denotes all the pixels that show parts of the surface in an \emph{indirect} way, and region $\cI^0$ denotes pixels that do not carry information about the surface, for example, they see the background clear color.
A (straight) view ray can be defined as $p_{\vi_\cD}(\lambda)=\vi_\cD+\lambda\vr_{\vi_\cD},\;\lambda\in\R^+$.
For all pixels $\vi\in\cI^+$ that hit a surface, the partial map $\cP$ is already defined:
\begin{align}\label{Eq:ProjectionWithCurve}
    \cP(\vi)=\argmin_{\vs\in\cS}\min_{\lambda\ge 0} \Vert p_{\vi_\cD}(\lambda)-\vs\Vert.
\end{align}
If only a projective mapping is used, then $\cI^-$ is empty.
Our goal is to define the image region $\cI^-$ and to construct a visual mapping $\cP$ for $\cI^-$ that conveys information about occluded surface parts.

\paragraph*{Map-based Visualization.}
Neugebauer et al.~\cite{Neugebauer09} constructed a projection from a center $\vc\in\cD$ in the scene.
They defined $\cI^-$ as 
$\cI^- = \{ \vi\in\cI \;|\; r_1 \le \|\vi - \cP^{-1}(\vc)\| \le r_2 \}$
where $r_1,r_2$ define the inner and outer circle radius and $\cP^{-1}(\vc)$ is the back projection of the center to the screen plane. 
The 3D coordinates of $\cP^{-1}(\vc)$ will be denoted as $\vc_\cD$.
For every $\vi\in\cI^-$, an angle $\alpha=\arctan(\vi)$ and a radius $r_\vi=\|\vi\|$ can be determined.
A (straight) ray is then defined as
\begin{align}
     p_{\vi}(\lambda)=\vc_\cD+\lambda\cdot\left(\frac{\vi-\vc_\cD}{\|\vi-\vc_\cD\|}+\left(\frac{2r_\vi}{r_1+r_2}-1\right)\cdot\vn\right),\;\lambda\in\R^+
\end{align}
where $\vn$ is the normal vector of the near plane $\cI_\cD$. 
Similarly, one can also determine the backsides of the surface and display them on the left-hand side of the screen. 
Again $\cP$ is defined as in Eq.~\eqref{Eq:ProjectionWithCurve}.
Their approach can be seen in Fig.~\ref{fig:teaser} (left).

\paragraph*{Mirror.} An alternative is to place a mirror behind the scene.
Let $\cM \in \cD$ be a (possibly curved) mirror surface with the normal field $\hat\vn(\vx)$.
For $\vi\in\cI^-$, let $\vi_\cM$ be the intersection point of $\vi_\cD+\lambda\cdot\vr_{\vi_\cD},\;\lambda\in\R^+$ with $\cM$.
Then, a (straight) ray is defined: 
\begin{align}
     p_{\vi}(\lambda)=\vi_\cM+\lambda\cdot\hat\vn(\vi_\cM),\;\lambda\in\R^+,
\end{align}
and $\cP$ is defined as in Eq.~\eqref{Eq:ProjectionWithCurve}, if the distance of the curve to the surface is equal to 0.
This approach is shown in Fig.~\ref{fig:teaser} (middle).

\begin{figure}[t]%
{
    \centering%
    \includegraphics[width=0.32\linewidth]{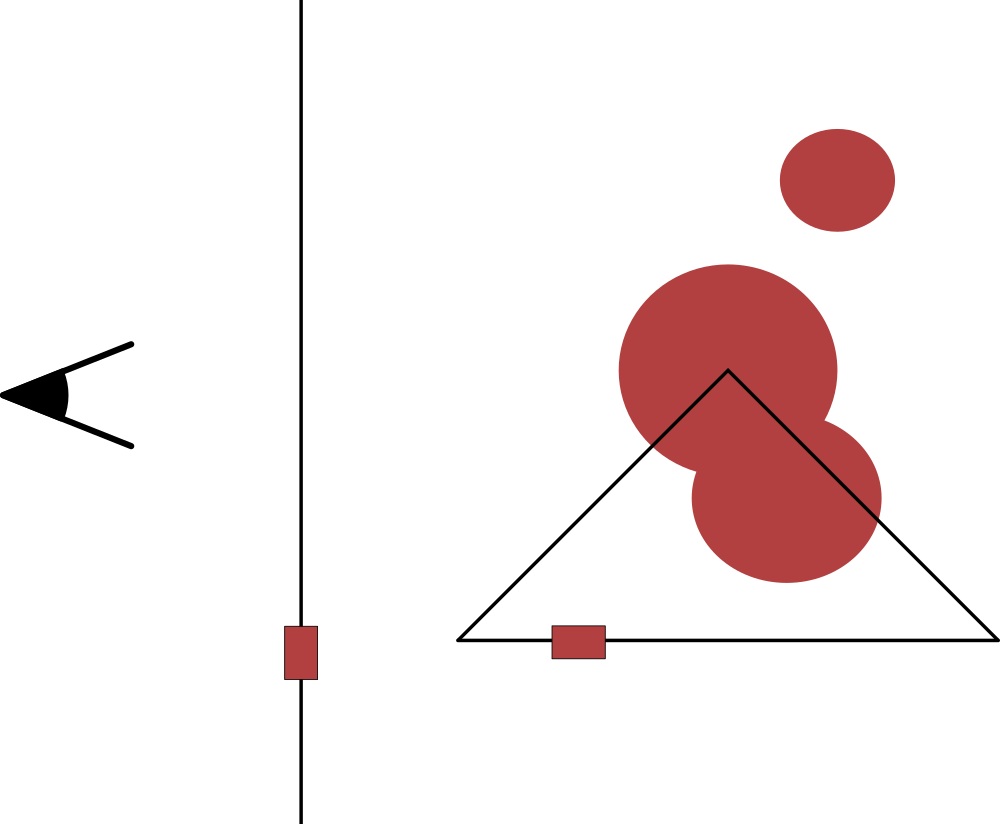}\hfill%
    \includegraphics[width=0.32\linewidth]{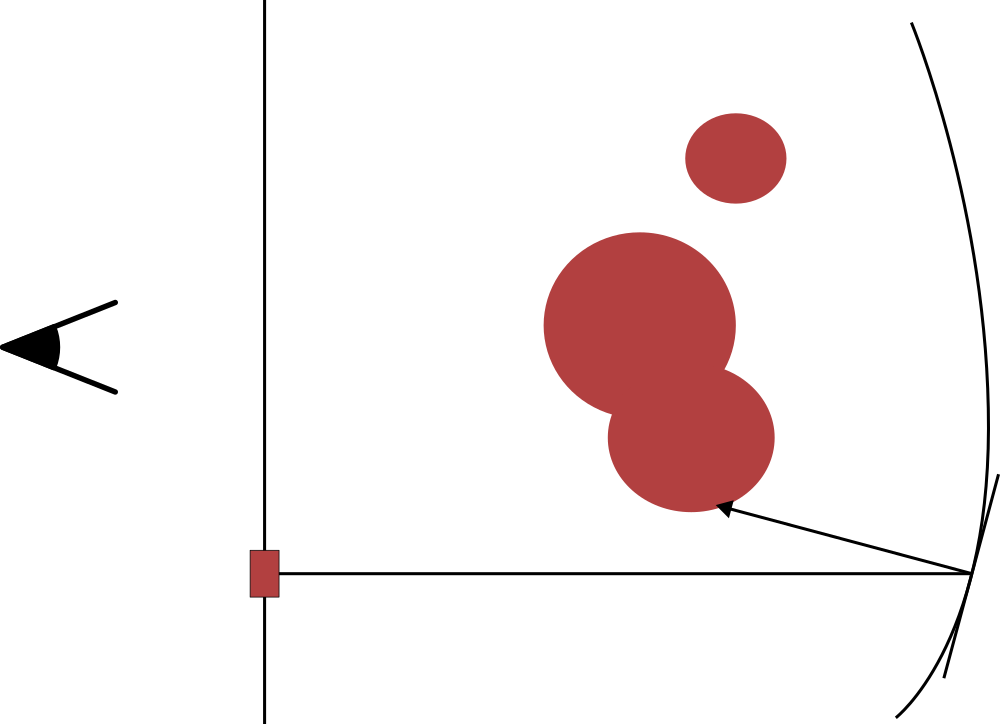}\hfill%
    \includegraphics[width=0.32\linewidth]{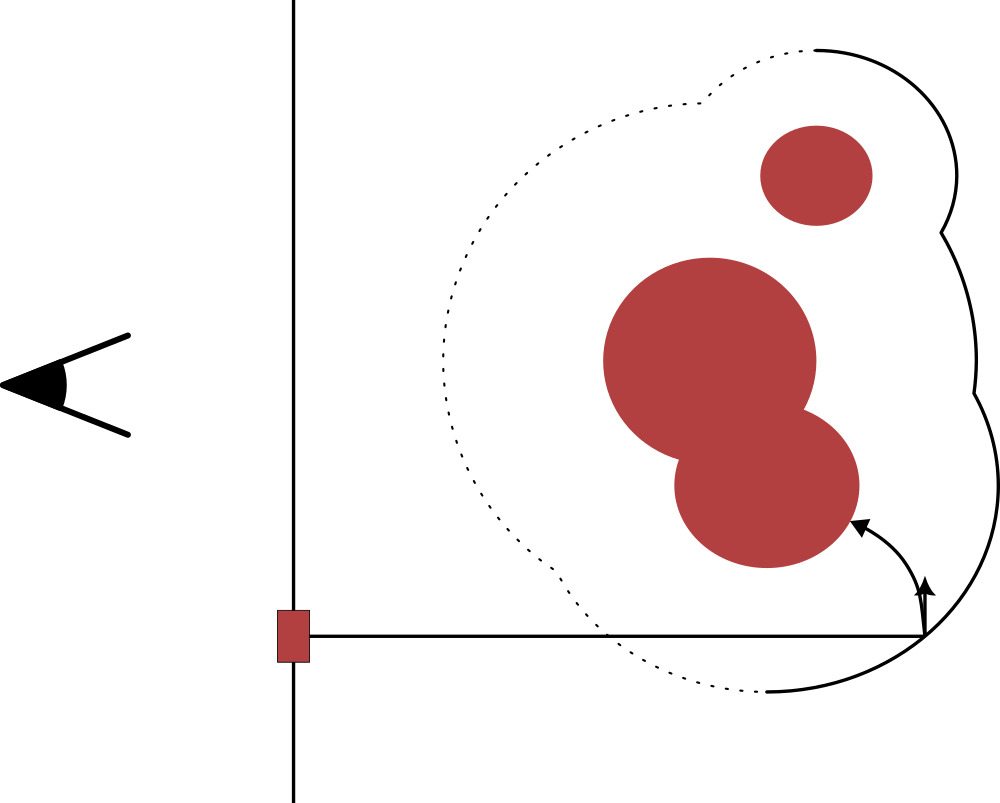}\\%
}
    \hspace{2.2cm}
    (a) Neugebauer et al.
    \hspace{2.1cm}
    (b) Optimized mirror
    \hspace{2.8cm}
    (c) InverseVis
    \caption{Schematic illustrations: (a) map-based visualizations by Neugebauer et al.~\cite{Neugebauer09}, (b) optimized mirror, (c) InverseVis.}%
    \label{fig:invervis-illustration}%
\end{figure}%

\section{Method}
Previous surface visualization methods had difficulties visualizing scalar fields on surface parts that are occluded or are facing away from the camera.
In this paper, we propose a new visual encoding of hidden surface geometry on parts of the screen that would otherwise have been unused (Sec.~\ref{sec:curved-ray}).
Further, we propose a viewpoint selection algorithm that determines the view that conveys the most information (Sec.~\ref{sec:viewpoint-optimization}).
To judge the quality of the resulting visualization, we introduce an energy that measures how well the relevant parts of the surfaces are shown.
For this, let $s(\vx): \cS \rightarrow [0,1]$ be a non-negative importance function that expresses how relevant each part of the surface is to see, where $0$ means irrelevant and $1$ means highly relevant.
Then, the energy to maximize is:
\begin{align}\label{Eq:Energy}
    E[\cP] = \gamma \underbrace{\int_{\cI^+} s(\cP(\vi)) \; \diff \vi}_{\textrm{visible part}}
      \;+\;  \underbrace{\int_{\cI^-} s(\cP(\vi)) \; \diff \vi}_{\textrm{occluded part}}
\end{align}
where the first term accounts for the information seen by regular surface rendering, and where the second term measures the information revealed by our proposed visual encoding.
The optional weight $\gamma$ allows favoring direct visibility, which we kept at $\gamma=1$ in the paper.

\subsection{Curved Ray Formulation}
\label{sec:curved-ray}
Information on the backside of a surface cannot be shown directly under regular perspective projection because it is hidden by other geometry.
This opens a visual design degree of freedom: we may choose a spatial arrangement for the hidden information and place it elsewhere on the screen.
To maintain a good spatial scene understanding, we aim to show the hidden information in an empty screen space that is \emph{close} to where the surface would normally be.
Since the hidden information is on the backside of the surface geometry, we place a \emph{new} geometry behind the surface geometry, onto which we map the hidden scalar field.
We refer to this new surface as the \emph{hull surface}.
To determine a mapping from the new geometry to the hidden surface, we propose a physically inspired model: for each point on the new geometry, we let a particle fall in a gravitational field onto the hidden surface.
This results in non-linear rays that can reach important regions on the backside, while establishing a well-behaved mapping in the sense that nearby points remain close.
The concept is illustrated in Fig.~\ref{fig:invervis-illustration}.
In the following, we explain the necessary ingredients in more detail: the construction of the hull geometry, the seeding of the particles, the governing equation that describes their motion, and an optimization of the model parameters to maximize the visibility of the surface.

\paragraph*{Hull Surface.}
The new hull surface onto which we aim to map the scalar field should be close to the original surface $\cS$.
For this purpose, we describe it as a level set in the signed distance field of the surface $\cS$.
Thus, we first convert the scene into an implicit representation.
Let $\phi(\vx) : \cD \rightarrow \mathbb{R}$ be the signed distance field to $\cS$, i.e., $\|\nabla\phi(\vx)\|=1$ and $\phi(\vx)=0$ for all $\vx\in\cS$.
The distances are positive on the inside and negative on the outside.
We then define the \emph{hull surface} $\cH \subset \cD$ as isosurface of $\phi(\vx)$ with isovalue $\phi_0$:
\begin{align}
    \cH = \{ \vx \;|\; \phi(\vx) = \phi_0\}.
\end{align}
The hull surface $\cH$ encloses the original surface $\cS$ entirely.
For each fragment, we use only the part of the hull surface that is farthest from the camera.
Formally, for all $\vi\in\cI^-$, let $p_{\vi}(\lambda)=\vi_\cD+\lambda\cdot\vr_{\vi_\cD},\;\lambda\in\R^+,$ be a projective mapping, but let $\vi_\cH$ be the intersection point of $p_{\vi}$ with $\cH$, such that $\vi_\cH=p_{\vi}(\lambda_{max})$, $\lambda_{max}=\argmax_{\lambda\in\R^+}\{\lambda\in\R^+\;|\;\|p_{\vi}(\lambda)-h\|=0,\;h\in\cH\}$.
Thus, it is the intersection point of the projective mapping with the hull surface, which is farthest away from the camera.
The point $\vi_\cH$ is located behind the object and is treated as a seed point for a curved ray towards the surfaces.

\begin{figure}[t]%
{
    \centering%
    \includegraphics[width=0.4\linewidth]{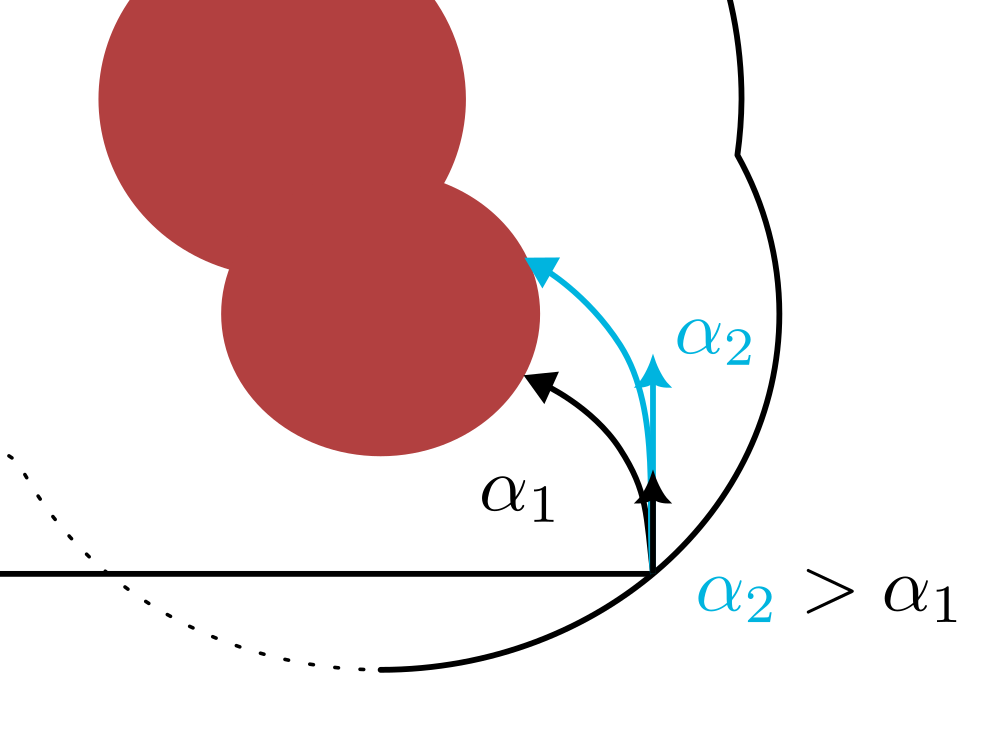}\quad%
    \includegraphics[width=0.4\linewidth]{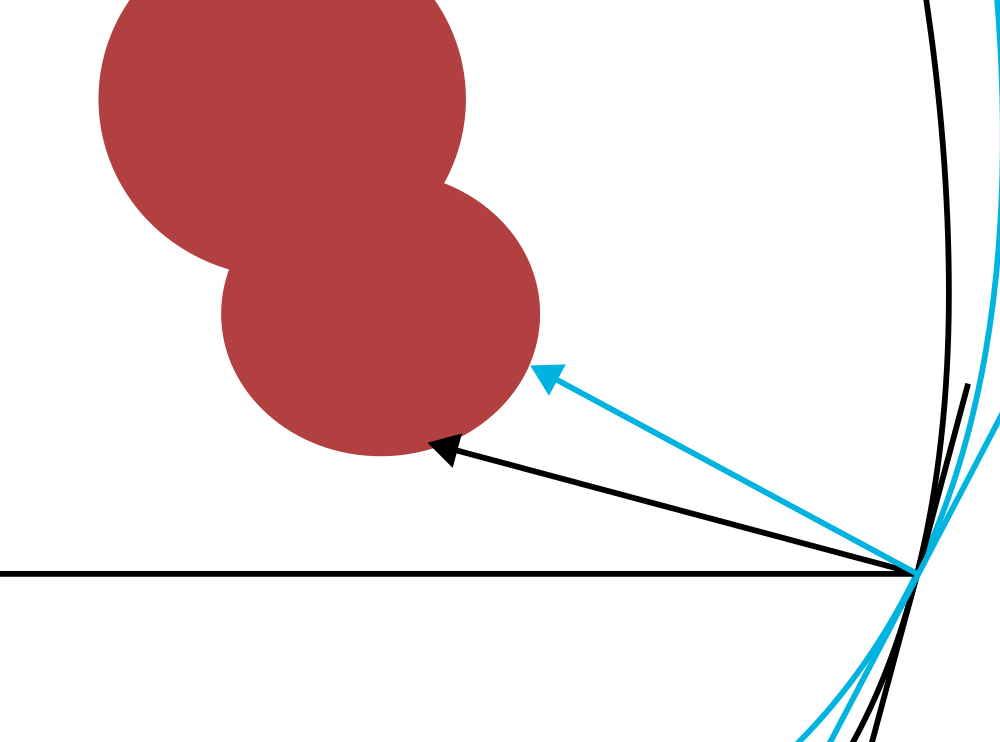}\\%
}
    \hspace{2.9cm}%
    (a) InverseVis
    \hspace{4.5cm}%
    (b) Optimized mirror
    \caption{InverseVis (a) optimizes a scale factor $\alpha$ of the initial velocity, while the mirror (b) is optimized for its surface normal.}%
    \label{fig:concept-alpha}%
\end{figure}

\paragraph*{Curved Particle Tracing.}
Intuitively speaking, we model the path from the hull surface onto the hidden surface by a particle that falls from the hull surface gravitationally onto the hidden surface of $\cS$.
As in Falk et al.~\cite{Falk07:Panorama},
the movement of a particle, which represents the ray, can be described by a velocity and an acceleration, which are formulated as an ordinary differential equation (ODE).
To model the path of a particle, we initially utilize sphere tracing, followed by ensuring that it descends onto the surface denoted by $\cS$.
The concept of sphere tracing involves selecting the particle's step size so that it advances in a direction equal to the shortest distance to the surface.
This allows taking larger steps compared to ray marching with a small step size.
Ultimately, this implies that the particle's velocity is determined by multiplying its current normalized velocity by the nearest distance to the surface.
The second principle dictates that the particle should be drawn towards the surface $\cS$, signifying in physics that the acceleration vector is directed toward the surface.
Both constraints can be described by a coupled first-order ODE:
\begin{align}\label{eq:path-integration}
    \frac{\diff}{\diff t}\vp&=\phi(\vp)\frac{\vv}{\|\vv\|}, &
    \frac{\diff}{\diff t}\vv&=-\nabla\phi(\vp).
\end{align}
Given initial conditions $\vp(0) = \vp_0=\vi_\cH$ and $\dot\vp(0) = \vv_0=\alpha\cdot\nabla\phi(\vi_\cH)\times(\vr_{\vi_\cD}\times \nabla\phi(\vi_\cH))$,
we numerically solve this ODE using the symplectic Euler method.
Again $\cP$ is defined as in Eq.~\eqref{Eq:ProjectionWithCurve}, if the distance of the curve $\vp(t)$ to the surface is equal to 0.
Note, that the parameter $\alpha$ controls the initial velocity, as shown in Fig.~\ref{fig:concept-alpha} (a), which has an impact on where the curve will hit the surface.

\paragraph*{Estimating the Impact of $\alpha$.}
Our goal is to optimize $\alpha$ to maximize the amount of information visible on screen, according to Eq.~\eqref{Eq:Energy}.
How much $\alpha$ impacts the trajectory can be studied through a continuum mechanical view onto a small variation of $\alpha$~\cite{Kasten2009LocalizedFTLE}.
For this, we first lift the coupled first-order ODE into a six-dimensional state $\bar\vx$, which changes in direction $\vF(\bar\vx)$:
\begin{align}
    \frac{\diff}{\diff t}\underbrace{\binom{\vp}{\vv}}_{\bar\vx}&=\underbrace{\binom{\phi(\vp)\frac{\vv}{\|\vv\|}}{-\nabla\phi(\vp)}}_{\vF(\bar\vx)}.
    \label{eq:phase-flow}
\end{align}
To study how a change in the initial condition of the ODE $\bar\vx = \vF(\bar\vx)$ carries over time, we add a small perturbation $\vdelta$:
\begin{align}
    \frac{\diff}{\diff t}(\bar\vx+\vdelta)&=\vF(\bar\vx+\vdelta)\\
    \frac{\diff}{\diff t} \bar\vx+\frac{\diff}{\diff t}\vdelta&\approx \vF(\bar\vx)+\mJ_\vF(\bar\vx)\cdot\vdelta\label{Eq:Taylor}\\
    \dot\vdelta&\approx \mJ_\vF(\bar\vx)\cdot\vdelta\label{Eq:1stOrderODE}
\end{align}
where we apply in Eq.~\eqref{Eq:Taylor} a first-order Taylor expansion around $\bar\vx$ to describe the evolution from information at $\bar\vx$.
Note that the change $\dot\vdelta$ is governed by the Jacobian $\mJ_\vF(\bar\vx)$ of the phase flow $\vF(\bar\vx)$:
\begin{align}
\mJ_\vF(\bar\vx) = \frac{\partial \vF(\bar\vx)}{\partial \bar\vx} = \begin{pmatrix}
        \frac{\vv}{\|\vv\|}\cdot\nabla\phi(\vp)^{\mathrm{T}} & \phi(\vp)\left(\frac{\mI_{3\times 3}}{\|\vv\|}-\frac{\vv\vv^{\mathrm{T}}}{\|\vv\|^3}\right) \\
        -\mH_\phi(\vp) & \vNull_{3\times 3}
    \end{pmatrix}.
    \label{eq:Jacobian}
\end{align}
Over time, the perturbation in Eq.~\eqref{Eq:1stOrderODE} is integrated by:
\begin{align}
    \vdelta(t) = \vdelta(0) + \int_0^t \underbrace{\mJ_\vF(\bar\vx(\tau))}_{\mJ(\tau)} \cdot \vdelta(\tau) \, \diff \tau
    \label{eq:perturbation-integral}
\end{align}
The integrand of Eq.~\eqref{eq:perturbation-integral} is a linear vector field in $\vdelta(\tau)$. 
Upon temporal discretization of this integral, the integration in the linear vector field can be solved analytically at each time step, integrating $\vdelta(t_{i})$ in the linear vector field up until the next time step $\vdelta(t_{i+1})$.
\begin{align}
   \vdelta(t_{i+1}) = e^{\mJ(t_i) h} \vdelta(t_{i})
   \label{eq:exp-recursion}
\end{align}
where $h = t_{i+1} - t_i$ is the numerical integration step size.
The matrix exponential $e^{\mJ(t) h}$ is calculated by diagonalization:
\begin{align}
    e^{\mJ(t) h} = \bar\mE(t) \cdot \bar\mD(t,h) \cdot \bar\mE(t)^{-1}
\end{align}
where $\bar\mE(t)$ is the eigenvector matrix of $\mJ(t)$, and where $\mD$ is the diagonal matrix of eigenvalues of $\mJ(t)$:
\begin{align}
    \bar\mE(t) = \begin{pmatrix}
        \bar \vc_1, \dots, \bar \vc_6
    \end{pmatrix}, ~~~\mD(t) = \mathrm{diag} \begin{pmatrix}
        \bar \lambda_1, \dots, \bar \lambda_6
    \end{pmatrix}
\end{align}
and where $\bar\mD(t,h)=e^{\mD(t) h} = \mathrm{diag}\begin{pmatrix}
        e^{\bar\lambda_1(t)\;h}, \dots, e^{\bar\lambda_6(t)\;h}
    \end{pmatrix}$ is the matrix exponential of the eigenvalue matrix computed at time $t$ for step size $h$.
Concatenating Eq.~\eqref{eq:exp-recursion} over all time steps, this gives:
\begin{align}
   \vdelta(t_{n}) = \underbrace{\left(\prod_{i=n-1}^0 e^{\mJ(t_i) h}\right)}_{\mathbf{\psi(t_n)}} \vdelta(0)
   \approx
   \underbrace{\left(\prod_{i=n-1}^0 \mI + h \mJ(t_i)\right)}_{\textrm{1st-order Taylor approx.}} \vdelta(0)
   \label{eq:recursive-perturbation}
\end{align}
which can be first-order Taylor approximated.
The matrix $\mathbf{\psi}_{6\times 6}$ tells how an initial perturbation $\vdelta(0)$ impacts the end of the path integration in Eq.~\eqref{eq:path-integration}. Since $\alpha$ is a linear scaling of the velocity subspace of the initial condition $\vv_0$, its impact is:
\begin{align}
    \frac{\partial\bar\vx(t)}{\partial \alpha} = \mathbf{\psi}(t) \cdot \left(\mathbf{0}_3, \frac{\partial \vv_0}{\partial \alpha}\right)^{\mathrm{T}}.
    \label{eq:alpha-partial}
\end{align}
The first three components denote the change in the position $\frac{\partial\vp(t)}{\partial \alpha}$.
With $t$ being chosen to be the time of the surface hit, this gives $\frac{\partial \cP}{\partial \alpha}$.

\paragraph*{Optimization.}
Knowing from Eq.~\eqref{eq:alpha-partial} the impact that a variation of $\alpha$ has, we can perform a gradient-based optimization of $\alpha$ to find the setting that maximizes the energy $E$ in Eq.~\eqref{Eq:Energy} best:
\begin{align}\label{Eq:GradientAscent-alpha}
    \alpha^{(i+1)}=\alpha^{(i)}+h\cdot \frac{\partial E}{\partial \cP} \frac{\partial \cP}{\partial \alpha},
\end{align}
where $\alpha^{(0)}$ is an initial guess, which we set to $0.5$, and where the energy partial with respect to the endpoint is $\frac{\partial E}{\partial \cP} = \nabla s(\cP)$.
During user interaction, a gradient-based optimization is favorable over gradient-free methods, since the optimized $\alpha$ adjusts smoothly over time.
To accelerate the convergence, an optimal step size $h$ can be determined by a line search, for which we employ a golden-section search~\cite{KieferFibonacciSearch1953} within the empirically chosen interval $h\in [0, 0.25]$. 
When given the interval $[X_1,X_4]$, it is divided in four parts: $X_1,X_2,X_3,X_4$, with $X_2=X_4-(X_4-X_1)/\Phi$ and $X_3=X_1+(X_4-X_1)/\Phi$, where $\Phi=\frac{1+\sqrt{5}}{2}$ is the golden ratio.
If $E_i+X_2\nabla E_i > E_i+X_3\nabla E_i$, then we repeat this procedure with the new interval $[X_1,X_3]$ otherwise, we use the interval $[X_2,X_4]$.
This is repeated until the length $L([a,b])=b-a$ of the interval falls below a threshold (in our case,  $0.01$), at which point $h$ is set to the interval's mean value.
After a few iterations, we identify the optimal step size $h$, and the gradient ascent is continued. 
The iteration halts if the energy decreases after the line search compared to its value before the search.

\subsection{Viewpoint Optimization}
\label{sec:viewpoint-optimization}
The previous section introduced a visual encoding that conveys more information about hidden surface geometry.
Since the approach is maximizing the energy in Eq.~\eqref{Eq:Energy} for a given viewpoint, the interesting question arises, to whether an even better viewpoint could be found automatically.
Previous research has investigated various methods for determining optimal camera positions, e.g. those that maximize high values in the scalar field~\cite{Meuschke_2017_BVM} or enhance visibility~\cite{Vazquez_2001}.
In our work, we maximize the energy in Eq.~\eqref{Eq:Energy} for every camera position to find the best view.

During our experiments, while searching for camera points that would maximize the energy, we encountered numerous local maxima. 
This necessitated the implementation of an optimization method capable of locating a global optimum within the camera space.
To address this challenge, we used \emph{simulated annealing}, which proved to be satisfactory both in terms of results and performance.
In our configuration, the camera is permitted to move along a sphere that encircles the object in focus. 
Without loss of generality, the object's dimensions are confined within $\vp\in[-1,1]^3$, and the radius of the surrounding sphere is fixed at $2.5$, which ensures that the whole object is visible and a margin exists that can be used for surrounding visualization.
Consequently, the camera movement is defined by two degrees of freedom, specifically the polar and azimuthal angles, denoted as $\theta$ and $\varphi$, respectively.

\begin{figure}[t]%
    \centering%
    \includegraphics[width=0.4\linewidth]{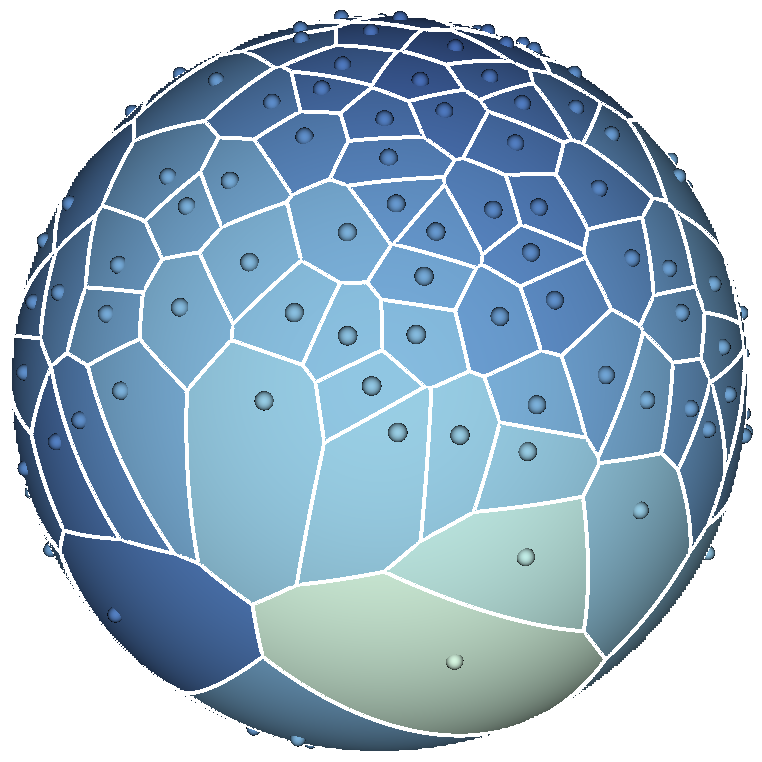}\hspace{2em}%
    \includegraphics[width=0.4\linewidth]{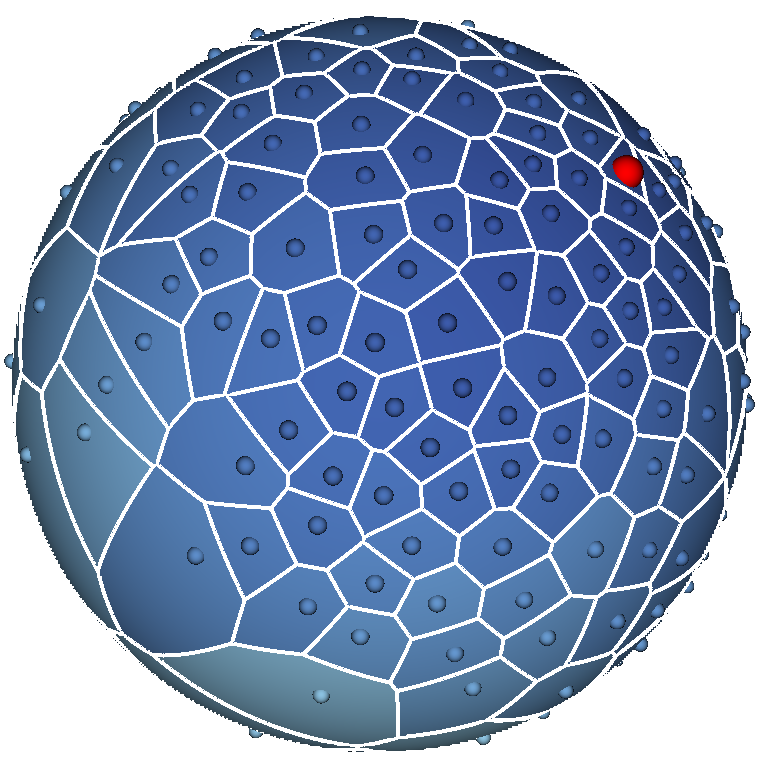}%
    \caption{This sphere shows the tested camera positions during the simulated annealing (front and backside). Dark blue represents values of high energy, which have a higher sample rate compared to less optimal regions. The found optimum is shown in red.}%
    \label{fig:SimAnnSphere}%
\end{figure}

Simulated annealing begins by determining the energy based on the camera's current position. Subsequently, a new camera position is randomly selected within the neighborhood of the existing position.
The range for this neighborhood is set to $\pm 60^\circ$. 
We found that this range is optimal as smaller angles tend to confine the search to local areas, while larger angles might prevent the algorithm from refining the optimum on a local scale.
When comparing the energy of the new camera position against the old one, a position is retained if yields higher energy. 
However, if the new position has less energy, it is still accepted with a probability of $\exp\left(10\cdot\frac{E_{new}-E_{old}}{T}\right)$, where $T$ represents the temperature.
The temperature starts at $1$ and gradually decreases with each step. 
Illustrated in Fig.~\ref{fig:SimAnnSphere}, we can observe the regions on a sphere visited by the camera during the simulated annealing process. In this instance, using the aneurysm dataset, we set $s$ in Eq.~\eqref{Eq:Energy} to represent visibility. This process showcases the effectiveness and adaptability of our simulated annealing approach in navigating and optimizing camera positions for complex surfaces.

\section{Implementation}
\subsection{Signed Distance Field}\label{Sec:SDF}
At its core, our objective is to trace rays around a surface to visualize its backside.
Since the rays are non-linear a ray-triangle intersection test after each numerical integration step is computationally expensive.
To expedite this process, we integrate rasterization and ray marching. 
For rendering the front faces of the surface, we utilize rasterization as facilitated by the graphics pipeline. 
For linear rays that encompass the surface, such as for casting shadow rays, we employ ray marching.
To accelerate this, we utilize sphere tracing, which requires a signed distance field (SDF).
We construct a voxelized space around this surface, which ranges in $[-2.5,2.5]$.
Inside this box, we construct $200$ voxels in every dimension.
The higher the number of voxels, the higher the resolution for the rendered backside of the surface.
We found that $200$ gives satisfactory results.
For every voxel, we determine the closest point on the surface and store: (i) the distance, (ii) the corresponding triangle ID, and (iii) the barycentric coordinates of the closest point inside the triangle. 
During the sphere tracing process, we efficiently intersect the surface at a given distance. This not only facilitates a rapid computation but also provides us with the corresponding triangle and its barycentric coordinates. This crucial information enables us to effectively visualize the scalar field, as well.
Additionally, we upload the triangle vertex indices to the GPU, where they are accessible alongside other data (i)-(iii) in a shader storage buffer.
On the GPU, the available distances are accessible as a 3D texture, enabling us to perform sphere tracing on the surface. Through this process, we first identify the potential triangle ID that would be intersected, followed by obtaining the vertex IDs of the triangle. The barycentric coordinates then pinpoint the exact position, allowing us to accurately determine the scalar field at that specific location.

\subsection{Mirror Optimization}\label{Sec:MirrorOp}
We will compare our InverseVis method with an optimized curved mirror that is in the shape of a quadratic height field $H(\vx)$:
\begin{align}\label{Eq:Mirror}
    H(\vx)=z-\omega_1 x^2-\omega_2 y^2-\omega_3 xy-\omega_4 x-\omega_5 y=0
\end{align}
with $\vx=(x,y,z)$.
This height field is placed behind the surface at a distance of $1$,
which could also be optimized, resulting in six degrees of freedom. 
The normals $\vn(x,y)$ of this height field are:
\begin{align}
    \vn(x,y)=\frac{\partial H(\vx)}{\partial \vx} =\begin{pmatrix}
        -2\omega_1 x - \omega_3 y - \omega_4\\
        -2\omega_2 y - \omega_3 x - \omega_5\\
        1
    \end{pmatrix}.
\end{align}
We utilize this normal as the reflection vector to visualize the backsides.
On the GPU, we intersect the mirror at the point $\vm(x,y)=\vm_0$ and determine its normal $\vn(x,y)=\vn_0$.
Then, sphere tracing is employed to intersect the surface $\cS$ applying
the forward Euler method:
\begin{align}
    \vp_{i+1}&=\vp_i+\phi(\vp_i)\cdot\vn_0, ~~~~~
    \vp_0=\vm_0.    
\end{align}
The mirror has five degrees of freedoms $\omega_1,\ldots,\omega_5$, which we optimize to maximize the energy Eq.~\eqref{Eq:Energy}, see Fig.~\ref{fig:concept-alpha} (b).
Similar to the $\alpha$ optimization in Eq.~\eqref{Eq:GradientAscent-alpha}, we perform a gradient-based optimization with a linear search.
That is, we calculate the energy, then, for every $\omega_i$, we slightly alter the parameter by an $\epsilon=0.025$ to determine the resultant energy change, necessary for calculating $\frac{\partial E}{\partial \omega_i}$.

\subsection{Symplectic Euler for InverseVis}\label{Sec:SymplecticInverseVis}
Tracing the non-linear rays as governed by the ODE from Eq.~\eqref{eq:path-integration} involves using a numerical procedure.
For second-order ODEs, a symplectic Euler method shows better energy conservation compared to explicit integration methods:
\begin{align}
    \vp_{i+1}&=\vp_i + h\cdot \phi(\vp_i)\frac{\vv_{i+1}}{\|\vv_{i+1}\|}, &
    \vv_{i+1}&=\vv_i - h\cdot\nabla\phi(\vp_i).
\end{align}
In our implementation, we use $h=0.1$ as a step size.

\subsection{OpenGL Implementation}
\label{sec:OpenGLImplementation}
Next, we outline the implementation steps of our visualization techniques.
Initially, we determine the SDF for a given surface mesh, which includes distances, triangles, and barycentric coordinates (refer to Sec.~\ref{Sec:SDF}). 
Following this, we either identify the intersection point with the mirror or with the hull surface at a radius of $0.4$.
Depending on whether we use the mirror or InverseVis, two main tasks are required: (i) optimizing the corresponding parameters and (ii) implementing the forward Euler method (see Sec.~\ref{Sec:MirrorOp} and Sec.~\ref{Sec:SymplecticInverseVis}).

For each parameter under optimization, we increment it by a small value $\epsilon=0.025$ (i) and then trace the path to the surface (ii). 
The process for the mirror involves sphere tracing, whereas, for InverseVis, we employ the forward Euler method with a step size determined through the golden-section search (Sec.~\ref{sec:curved-ray}).
After adjusting every parameter with $\epsilon$, we calculate the gradient using forward differences and again use the golden-section search to find the optimal parameter combinations.

\begin{figure*}[t]%
{
    \centering%
    \includegraphics[width=\linewidth]{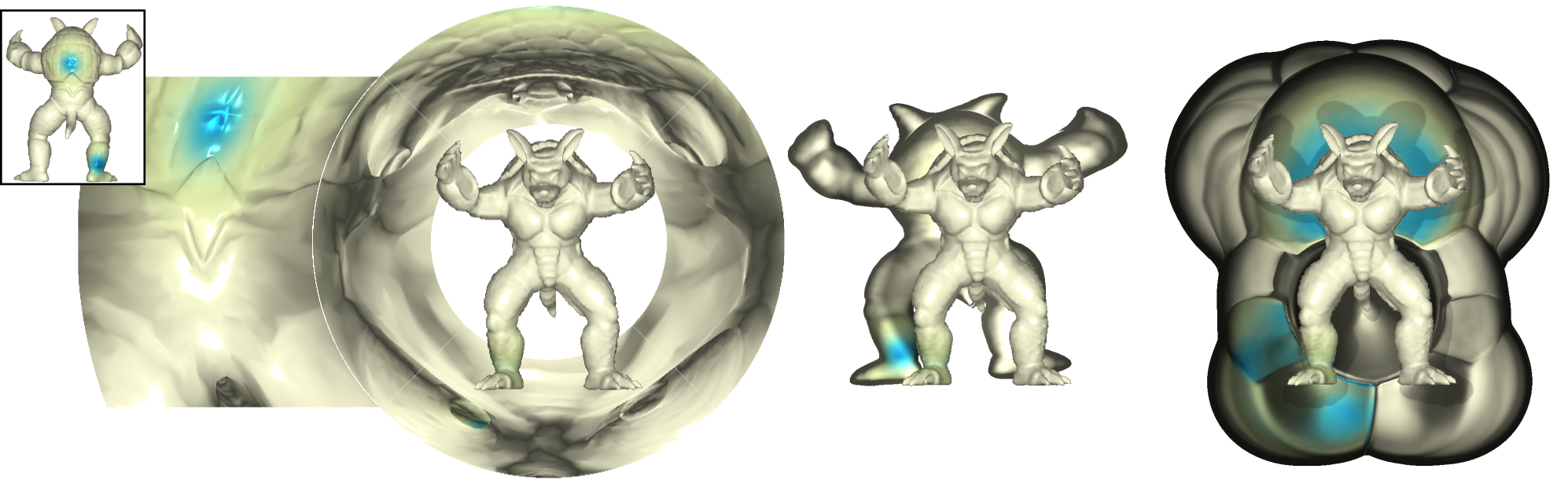}\\%
}
    \hspace{3.1cm}
    (a) Neugebauer et al.
    \hspace{2.2cm}
    (b) Optimized mirror
    \hspace{1.7cm}
    (c) InverseVis
    \caption{In this scene, two distinct hidden regions are marked as important (on the back and on the leg). Neugebauer et al.~\cite{Neugebauer09} shows the back, but the spot on the foot is rather unnoticeable. The mirror shows only the foot. InverseVis is able to reveal both hidden regions.}%
    \label{fig:ArtificialHotspots}%
\end{figure*}%

The energy, cf. Eq.~\eqref{Eq:Energy}, can be based on either the scalar field or visibility.
For the scalar field, we add the scalar values of each ray that hits the surface.  
This is achieved using the OpenGL extension \texttt{GL\_NV\_shader\_atomic\_float}, which facilitates the addition of float values in the fragment shader.
We also use an atomic counter to track the frequency of surface hits, which allows us to calculate the average scalar value in the image.
If optimizing for visibility, we create a 3D texture of size $200\times 200\times 200$ and mark all voxels within a distance smaller than the voxel size to the surface $\cS$, identifying voxels near the surface.
To assess visibility, we count voxels that are hit by rays. 
When a ray intersects a surface, we locate the current voxel and increment a count if it is a valid (marked) voxel that has not been visited by another ray.
We then calculate the ratio of all marked and visited voxels.
Optionally, users can determine the optimal camera position based on different energies (scalar field or visibility), cf Sec.~\ref{sec:viewpoint-optimization}.

\section{Evaluation}
To assess the effectiveness of our new methods, which are designed to reveal the backsides of surfaces, we carried out three evaluations.
Firstly, it was crucial to quantify how much of the surface is visible when employing these different methods. 
Achieving values close to 100\% indicates that the method is successful in achieving its goal of comprehensive surface visibility. 
The second aspect of our evaluation focused on the time required to identify optimal camera positions.
A technique, despite offering excellent visibility, might not be user-friendly if it requires an excessively long time for optimization.
Finally, the last part of our evaluation involved gathering qualitative feedback from various visualization experts, providing insights into the practicality and efficiency of the different methods.

\subsection{Visibility Measure}
We focused on measuring visibility, or more precisely, the extent of the surface $\cS$ that is visible when using different techniques.
For this purpose, we utilized the voxelized space of the signed distance field, as detailed in Sec.~\ref{Sec:SDF}.  
In this process, we marked every voxel within a distance smaller than the voxel size to the surface $\cS$, effectively identifying all voxels in proximity to the surface.

\begin{figure*}[t]%
{
    \centering%
    \includegraphics[width=\linewidth]{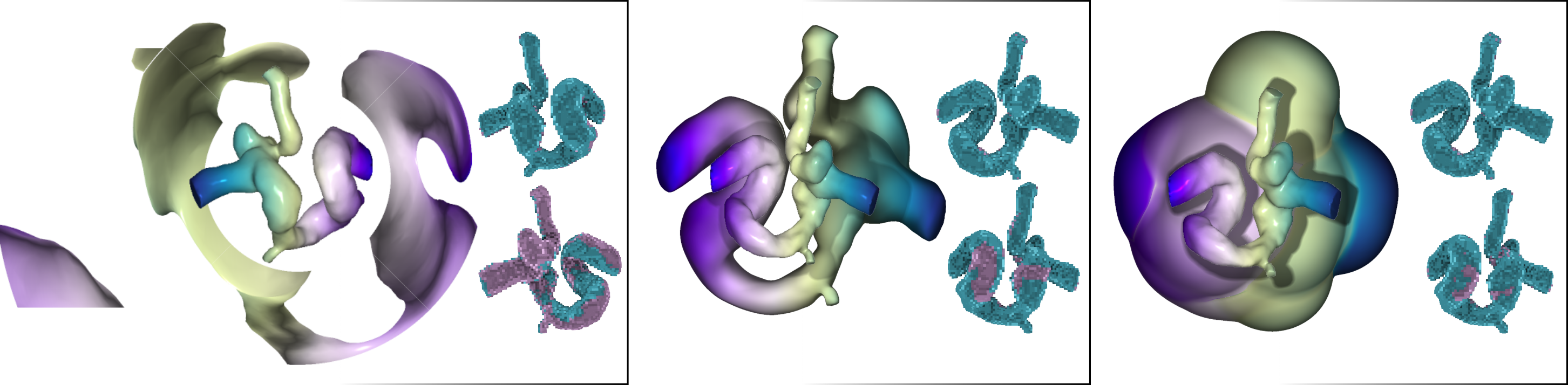}\\%
}
    \hspace{1.9cm}
    (a) Neugebauer et al.
    \hspace{2.5cm}
    (b) Optimized mirror
    \hspace{2.3cm}
    (c) InverseVis
    \caption{Optimal viewpoints on the Aneurysm 3 dataset. Shown are the techniques (vertices are colored according to their voxels) as well as the visible (teal) and non-visible (violet) voxels on the right with front and backside. From a total of 6,910 voxels, Neugebauer et al. ~\cite{Neugebauer09} reaches 4,518, the optimized mirror 6,173, and InverseVis 6,654.}%
    \label{fig:voxel_visibility}%
\end{figure*}

\begin{figure*}[t]%
{
    \centering%
    \includegraphics[width=\linewidth]{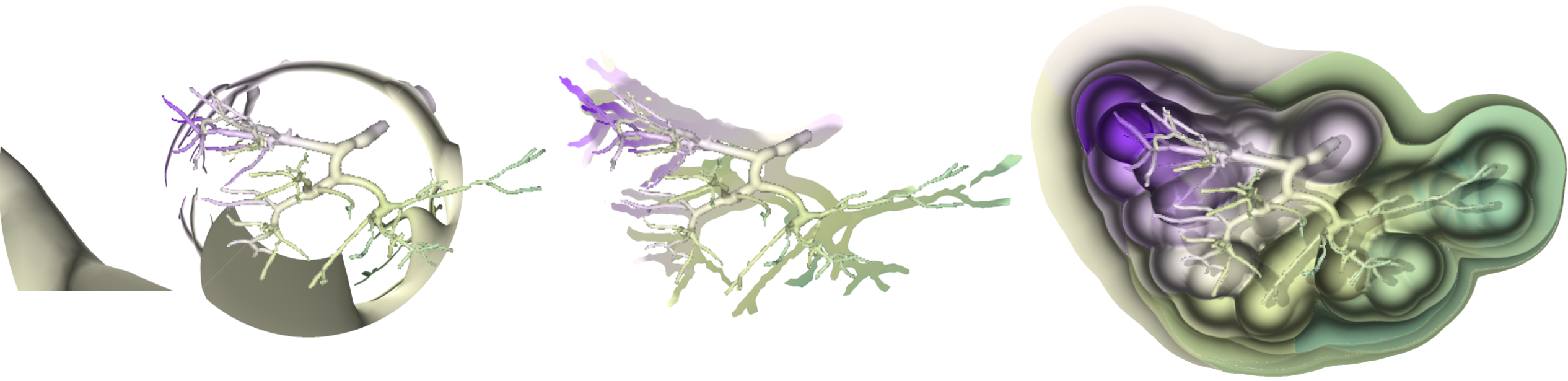}\\%
}
    \hspace{1.5cm}
    (a) Neugebauer et al.
    \hspace{1.6cm}
    (b) Optimized mirror
    \hspace{2.9cm}
    (c) InverseVis
    \caption{The vertices are colored according to their voxel, which is used for the visibility measure. Neugebauer et al. ~\cite{Neugebauer09} (left) captures only a small part of the backside due to the complexity of the surface and the reference point for the views, the optimized mirror (middle) reflects more of the backside but the space is not fully used, and InverseVis (right) uses cavities for tracing the rays to the backsides.}%
    \label{fig:Visibility}%
\end{figure*}

\begin{figure}[t]%
{
    \centering%
    \includegraphics[width=0.9\linewidth]{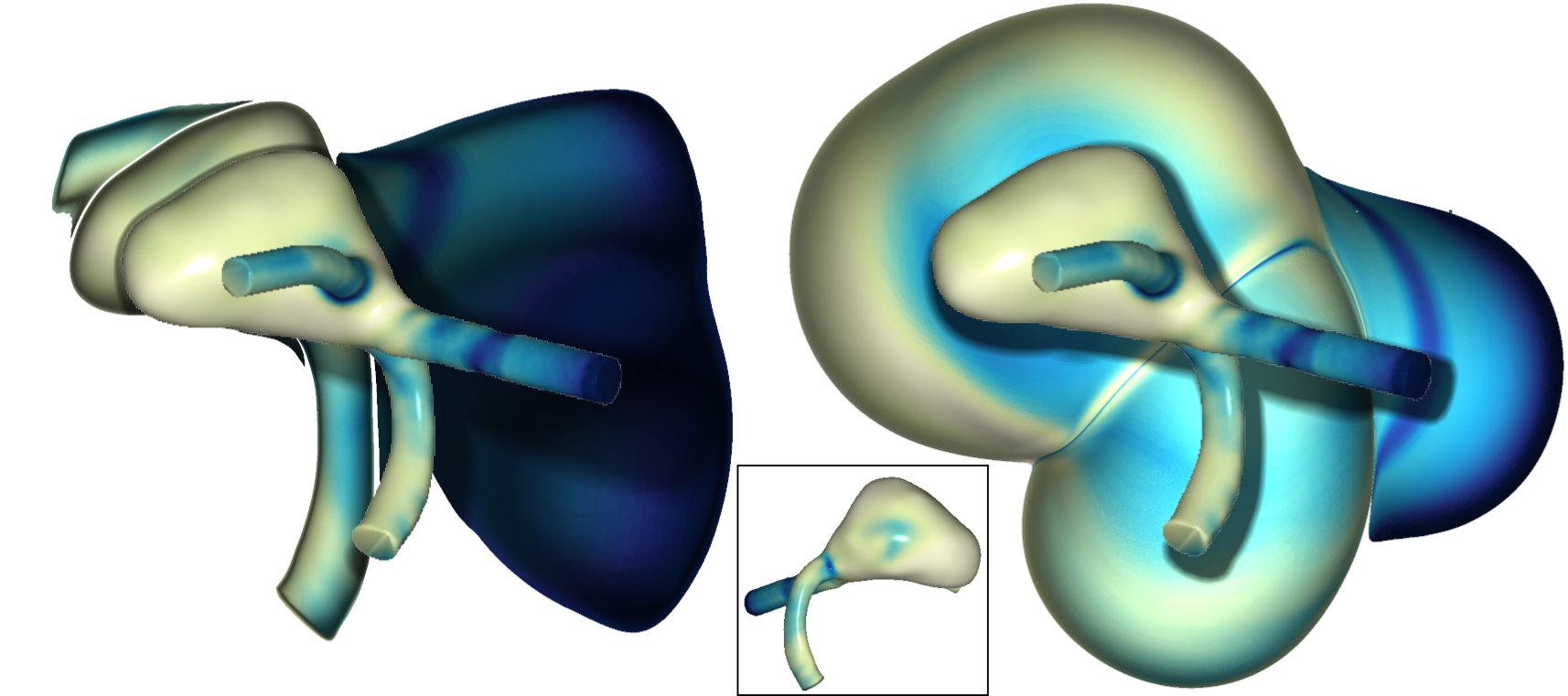}\\%
}
    \hspace{2.8cm}
    (a) Optimized mirror
    \hspace{4.4cm}
    (b) InverseVis
    \caption{The mirror optimization (left) significantly enlarges the right vessel area, which increases the energy visibility, yet overlooks other relevant areas of the aneurysm. In contrast, InverseVis (right) also shows the aneurysm's backside, ensuring a more balanced visual representation. The boxed image below displays the backside.}%
    \label{fig:InverseVisVsMirror_MI}%
\end{figure}

When camera rays intersect the surface, the corresponding voxels are designated as visited.
We count all visited voxels and divide this by the total number of marked voxels.
This calculation yields a value in $[0,1]$, indicating the portion of the surface that is visible.
To achieve the best results, we employed the simulated annealing approach described in Sec.~\ref{sec:viewpoint-optimization}, aiming to maximize visibility for each technique, as shown in Tab.~\ref{tab:Visibility}. 
Fig.~\ref{fig:ArtificialHotspots} shows a scene, in which two distinct regions on the surface are marked as important (blue).
Neugebauer et al.~\cite{Neugebauer09} and the optimized mirror, both fail to show both important structures, while our method clearly displays both regions.
Fig.~\ref{fig:voxel_visibility} and Fig.~\ref{fig:Visibility} illustrate the visibility achieved by all methods on different surfaces, with vertices colored according to their corresponding voxels used in the visibility measurement. 
The method by Neugebauer et al.~\cite{Neugebauer09} struggles to capture all backside information
due to the model's complexity, while the mirror reflects more of the backside but does not fully utilize the surrounding space. 
In contrast, InverseVis can capture the most information, showcasing its effectiveness in visualizing the surface.
Fig.~\ref{fig:InverseVisVsMirror_MI} gives an example of the mirror optimization focusing on only one small (yet important) part on the backside, which misses all other relevant structures on the backside, while the InverseVis approach can reveal all structures on the backside.


\setlength{\tabcolsep}{4pt}
\begin{table}
    \centering
    \caption{The visibility in percent of different surfaces compared to the methods. Percentages in brackets present the visibility of the front faces, rendered with direct rendering ($^\ast$are used for evaluation).}
    \begin{tabular}{l||c|c|c}
    \toprule
         Surface        & Neugebauer et al.  & Mirror & InverseVis \\\hline
         Armadillo$^\ast$ & 79\,\% (46\,\%)   & 88\,\% (47\,\%)   & \textbf{97}\,\% (45\,\%)  \\
         Cow$^\ast$       & 91\,\% (38\,\%)   & 95\,\% (38\,\%)   & \textbf{98}\,\% (47\,\%)  \\
         Fertility$^\ast$ & 71\,\% (41\,\%)   & 86\,\% (34\,\%)   & \textbf{96}\,\% (43\,\%)  \\
         Gargoyle$^\ast$   & 84\,\% (42\,\%)   & 88\,\% (39\,\%)   & \textbf{95}\,\% (37\,\%)\\
         Neptune$^\ast$   & 72\,\% (43\,\%)   & 95\,\% (53\,\%)   & \textbf{96}\,\% (53\,\%)  \\
         Rocker arm  & 84\,\% (55\,\%)   & 97\,\% (47\,\%)   & \textbf{98}\,\% (46\,\%)  \\
         Sphere         & 100\,\% (43\,\%)  & 75\,\% (43\,\%)   & \textbf{100}\,\% (43\,\%)  \\\hline
         Aneurysma 1$^\ast$    & 84\,\% (41\,\%)   & 92\,\% (43\,\%)   & \textbf{100}\,\% (43\,\%)  \\
         Aneurysma 2$^\ast$    & 81\,\% (41\,\%)   & 84\,\% (44\,\%)   & \textbf{96}\,\% (44\,\%) \\ 
         Aneurysma 3    & 65\,\% (37\,\%)   & 89\,\% (42\,\%)   & \textbf{96}\,\% (42\,\%)\\ 
         Carotis 5$^\ast$      & 78\,\% (36\,\%)   & 93\,\% (46\,\%)   & \textbf{98}\,\% (49\,\%)  \\
         Carotis 11$^\ast$     & 68\,\% (44\,\%)   & 97\,\% (45\,\%)   & \textbf{100}\,\% (49\,\%)  \\
         Vessel Tree$^\ast$     & 73\,\% (47\,\%)   & 84\,\% (48\,\%)   & \textbf{98}\,\% (45\,\%)  \\
         \bottomrule
    \end{tabular}
    \label{tab:Visibility}
\end{table}

\subsection{Timings}
In this section, we discuss the time required to identify an appropriate camera position that maximizes visibility.
It is important to note that the time taken to optimize the scalar fields closely mirrors that for enhancing visibility. 
Therefore, we will focus on reporting the latter, as detailed in Tab.~\ref{tab:timings}. 
Through empirical observation, we found that the camera consistently optimizes itself to the same position, regardless of whether the optimization resolution is set at $100$ or $1000$.
This observation suggests that choosing a lower resolution is advantageous, as it leads to a significantly quicker optimization process without compromising the optimal camera positioning.
The experiments were conducted on an Intel Core i9 @3.60GHz, 32 GB RAM, and an NVIDIA GeForce GTX 2080. 
\begin{table}
    \centering
    \caption{Timings in seconds of the different methods on various data sets. The display was rendered with a resolution of $1000\times 1000$. The optimization resolution is set to either $100$ or $1000$.}
    \begin{tabular}{l||cc|cc|cc}
    \toprule
         Surface        & \multicolumn{2}{c|}{Neugebauer et al.}  & \multicolumn{2}{c|}{Mirror} & \multicolumn{2}{c}{InverseVis} \\\hline
          Optim. Res.   & 100 & 1000 & 100 & 1000 & 100 & 1000 \\\hline
          Armadillo     & 10  & 37   & 18  & 300  & 13  & 40 \\
          Neptune       & 13  & 36   & 16  & 245  & 10  & 35 \\\hline
          Aneurysma 1   & 13  & 33   & 18  & 350  & 10  & 39 \\
          Carotis 5       & 23  & 42   & 26  & 298  & 23  & 120 \\
          Vesseltree    & 14  & 39   & 14  & 115  & 14  & 70 \\
           \bottomrule
    \end{tabular}
    \label{tab:timings}
\end{table}

\subsection{Qualitative Feedback}
\begin{figure}[t]%
    \centering%
    \includegraphics[width=0.8\linewidth]{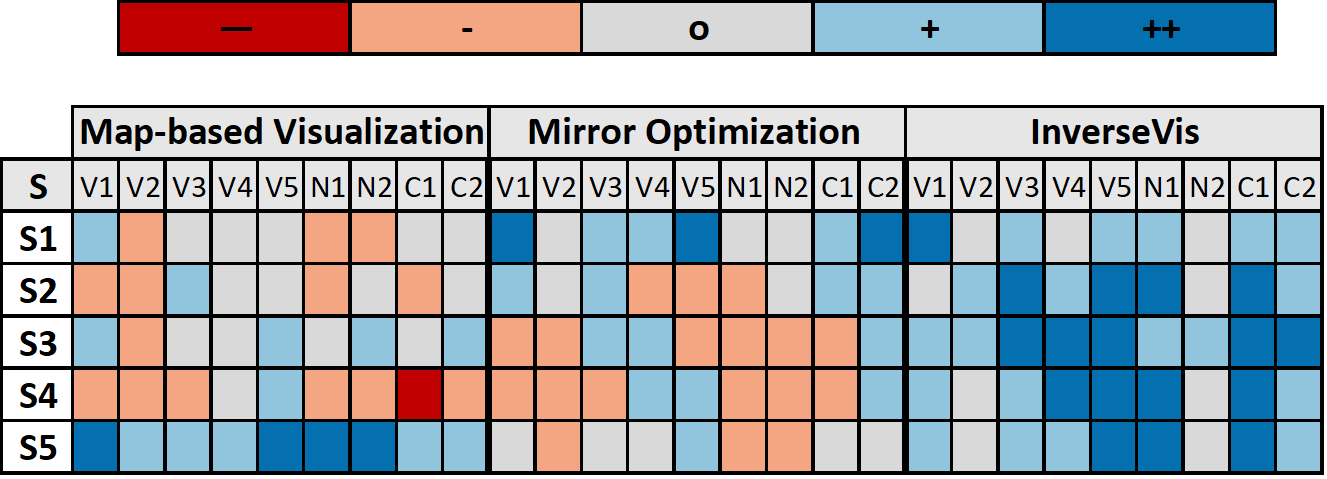}%
    \caption{Likert scores of the five study participants regarding the five statements for each of the techniques, Neugebauer et al. \cite{Neugebauer09} (left), optimized mirror (middle), and InverseVis (right).}%
    \label{fig:evalresults}%
\end{figure}

We conducted a qualitative user study with five visualization experts specialized in the visualization of biomedical data (V1-V5), two neuroradiologists (N1,N2)), and two CFD experts (C1,C2) to compare the method by Neugebauer et al.~\cite{Neugebauer09}, the \textit{mirror optimization}, and \textit{InverseVis}.
For this purpose, each expert explored six different datasets (a medical and a non-medical dataset for each method, see Tab.~\ref{tab:Visibility}, models are marked with $^\ast$). 
The study involves assessing the three methods regarding various aspects such as understandability, clarity in geometric correspondences, visibility of interesting regions, and performance with complex surfaces. 
All statements are structured to elicit responses on a 5-point Likert scale, ranging from 'Strongly Disagree' to 'Strongly Agree', see Fig.~\ref{fig:evalresults}. 
The statements (S1-S5) that each expert should evaluate for all three techniques are: 
\begin{itemize}
    \item \textbf{S1 Understandability:} This method is easy to understand.
    \item \textbf{S2 Geometric Correspondences:} It is easy to correlate scalar field values with their corresponding surface position. 
    \item \textbf{S3 Visibility of Interesting Regions:} This method effectively highlights the most interesting surface regions. 
    \item \textbf{S4 Handling of Complex Surfaces:} This approach performs well with surfaces of varying complexity.
    \item \textbf{S5 Clarity:} The visualization is both clear and well-organized.
  \end{itemize}

\noindent
\textbf{Results of the 2D Map}~\cite{Neugebauer09}. This approach faced some challenges in the user study, with eight experts rating it as less understandable (S1), indicating a general difficulty in grasping the method. Especially the medical experts found it difficult to understand that the back of the surface was projected onto a part to the left of the map ring. The experts quickly lost their orientation, especially when rotating more complex surfaces such as vascular trees. This difficulty further extended to correlating scalar fields with surface regions (S2), as reflected in the lower ratings from the experts. Due to the ring-shaped ordering of the projection results, the experts had problems correlating scalar field values with their corresponding surface position, especially in the case of complex surfaces, like vessel trees. A notable issue highlighted by most participants was the method's ineffectiveness in highlighting interesting regions (S3), suggesting a lack of clarity in visual emphasis.  Depending on the surface complexity and camera position, hotspots on the surface may be partially or completely hidden in the projection. For example, the neuroradiologists were interested in surface regions with high stress values in the aneurysm data sets. However, such regions are difficult to detect if they occur in concave regions at the junction with the parent vessel. This also explains the ratings for handling complex surfaces (S4), where most experts leaned towards the lower end of the scale. For complex surfaces with fine structures, such as vascular branchings, only very small sections of the surface can be projected. Despite the challenges associated with this method, experts appreciated its well-structured design in the 2D map format (S5). The medical experts, in particular, are very familiar with 2D representations, which is why they liked the general 2D concept of the map display. Particularly for convex surfaces without holes, the ring-shaped representation was regarded as clear and uncluttered. 

\noindent
\textbf{Results of the Mirror.} Six experts found the mirror method easy to understand (S1), noting that the concept of using a mirror is familiar and intuitive to most people. Three experts, in particular the neuroradiologists, struggled with its intuitiveness since, depending on the camera position and surface complexity, it might happen that surface parts are more deformed, leading to presentations that are difficult to understand. In terms of geometric correspondences (S2), participants were evenly divided, indicating a varied ability to correlate geometric features with the method. Again, possible strong distortions make it difficult to correlate scalar values and associated surface regions. The visibility of interesting regions (S3) was another area where the method received critique. Depending on the surface complexity and number of interesting regions, individual hotspots may be occasionally overshadowed. In line with previous assessments, handling complex surfaces (S4) proved challenging for some, with six experts finding it less effective in this regard, indicating difficulty in visualization. The deformation of the surface induced by the mirror can lead to confusing views, especially in the case of concave surfaces. This also explains the rather reserved assessment of its clarity (S5). This reasoning also sheds light on the somewhat cautious evaluation of its clarity (S5). The experts faced issues with orientation and navigation, particularly when adjusting the camera position and dealing with complex surfaces. 

\noindent
\textbf{Results of InverseVis.} InverseVis was received positively, with six experts rating it highly for its understandability (S1). Three experts indicated that it required some initial effort to learn, leading to neutral assessments. But they also mentioned that once they became accustomed to the method, it was easy to adapt to and use effectively. This positive response was echoed in its approach to geometric correspondences (S2), with the majority of experts (7 out of 9) assessing it as easy to correlate scalar field values with surface regions. Here, the experts also acknowledged the additional depth hints in the form of shadows that further support orientation. Only two experts encountered challenges with the correlation task, finding that the corresponding surface position was not always unambiguous, particularly when adjusting the camera position. In addition, InverseVis received very positive feedback regarding the visibility of interesting regions (S3), with all participants agreeing. Compared to the Mirror, InverseVis highlights intriguing regions on the surface, e.g., on the aneurysm compared to the less interesting inflow vessel, see Fig.~\ref{fig:InverseVisVsMirror_MI}. This shows the strength of the camera optimization to find interesting regions. Regarding the handling of complex surfaces (S4), the method received positive feedback, with a majority (7 out of 9) finding it effective, though two experts had some reservations. Specifically, for intricate structures like branched vessel trees, these experts experienced orientation difficulties with certain camera positions. This, in turn, also led to two cautious assessments regarding the structure of the presentation (S5), while all other experts perceived InverseVis as clear and well-structured. 

\noindent
\textbf{Final Ranking.} At the end of the user study, the participants were asked to rank the three methods regarding two major aspects:
\vspace{-0.2em}%
\begin{enumerate}
    \item Ease of Understanding
    \item Revealing Interesting Regions
\end{enumerate}
\vspace{-0.2em}%
The InverseVis method emerged as the clear leader in both ease of understanding and effectiveness in revealing interesting regions. Participants found the camera optimization to be the most intuitive and straightforward, attributing high scores to its understandability and ability to correlate geometric features with surface regions. Additionally, InverseVis was praised for its efficient highlighting of important areas without overshadowing or overemphasizing, and its adaptability to complex surfaces.
The Mirror ranked second in both categories, offering moderate ease of understanding. However, it faced challenges in consistently highlighting important regions and in its handling of complex surfaces. Its real-time deformation feature and adaptability were noted, but its balanced focus in visualization was questioned.
The Neugebauer et al. approach was found to be the least intuitive and understandable. Participants struggled with its learning curve, finding it challenging to correlate visualization aspects with surface regions and to highlight important regions effectively. It was also perceived as less efficient in handling complex surfaces, contributing to its lower ranking in these categories.

\section{Discussion}
The three methods aim to reveal hidden surfaces.
Neugebauer et al. use the arrangement of different rendered pages and the Mirror and InverseVis try to increase visibility through optimizations.
The mirror is slower than the other methods, which is because five parameters have to be optimized. 
For InverseVis and Neugebauer et al., the computation time and the comprehensibility both depend on the complexity of the mesh. 
InverseVis was slower for the vessel tree.
For simple convex objects, such as an aneurysm, Neugebauer et al.'s method is easy to understand once users have gained some practice. However, this was more difficult for the vessel tree, as gaps occur that lead to loss of orientation. 
This is reduced with the mirror, but as soon as it bends strongly to emphasize important regions, strong distortions occur, which we want to avoid.
These strong distortions cannot occur with Neugebauer et al. and InverseVis.  
We found that the camera position was not sensitive to the resolution used for optimization. 
We suspect that this was due to the large spatial extent of important structures, which were recognizable even at low resolution.
Certainly, the resolution will influence the data if there are small regions with high scalar values that the rasterizer does not cover. 
However, this generally leads to an interesting problem, which can also apply to normal direct rendering. 

The qualitative user study reveals insightful feedback from experts with varying backgrounds on the use of the different visualization methods. Despite their different levels of expertise, the experts did not show any significant differences in their assessments of the individual methods. Medical experts preferred 2D representations in principle but found the 2D map-based method challenging to understand, especially compared to the InverseVis method. Both medical and CFD experts required more time to familiarize themselves with the InverseVis method compared to visualization experts, suggesting the potential benefit of small tutorials for new users in the future. 

\section{Conclusions and Future Work}
Exploring surfaces with scalar fields can be a tedious process.
For this reason, we have addressed the issue of displaying the backsides in addition to the standard front rendering.
We took an existing method from Neugebauer et al., then improved an existing concept, the mirrors, and developed a new method, InverseVis.
We were able to optimize the new methods concerning energy so that either the visibility or interesting scalar fields are highlighted.
We tested the performance of the method by finding optimal camera positions and by comparing the visibility.
Finally, we evaluated the methods in a qualitative study with participants. 
The optimal mirror and InverseVis were able to significantly improve the previous method, as we were able to demonstrate with several examples and prove in our studies.
We are convinced that the combination of rasterization and ray marching, i.e., curved sphere tracing, offers a successful alternative and has great potential for the future.   

Currently, we have only dealt with static scalar fields. 
In the future, we want to deal with the question of how to find not only the best camera position but also good parameter combinations to then visualize as much information as possible on average over a time-varying scalar field.
Another interesting aspect would be to determine not only the (one) global optimum for the camera position but also the best viewpoints and to define a camera animation.
Furthermore, additional information related to the detected best viewpoints should be presented, e.g., interesting flow patterns close to vessel wall regions with highest shear forces~\cite{eulzer2021visualizing,neugebauer2013amnivis}.  
Besides, scalar fields, both vector and tensor fields mapped to surfaces play an important role, e.g., in analyzing medical flow simulation results~\cite{meuschke2017glyph}. Therefore, we plan to extend our optimization concerning the visibility of vector and tensor information.

\subsection*{Acknowledgements}
This work was supported by the Carl Zeiss Foundation, within the project Interactive Inference, and DFG grant no. GU 945/3-1.

\newpage

\newcommand{\etalchar}[1]{$^{#1}$}

\end{document}